\documentclass[12pt,a4,portrait]{article}

\usepackage{latexsym}
\usepackage{hyperref}
\usepackage{epsfig}
\newcommand{\be}{\begin{equation}}
\newcommand{\ee}{\end{equation}}
\newcommand{\beq}{\begin{eqnarray}}
\newcommand{\eeq}{\end{eqnarray}}
\newcommand{\bed}{\begin{displaymath}}
\newcommand{\eed}{\end{displaymath}}
\newcommand{\bc}{\begin{center}}
\newcommand{\ec}{\end{center}}
\newcommand{\bi}{\begin{itemize}}
\newcommand{\ei}{\end{itemize}}
\newcommand{\bn}{\begin{enumerate}}
\newcommand{\en}{\end{enumerate}}

\begin{document}

\title{Potential dominated scalar-tensor cosmologies\\in the general relativity limit:
phase space view}
\author{Laur J\"arv\thanks{laur.jarv@ut.ee}, Piret Kuusk\thanks{piret@fi.tartu.ee} \\
 Institute of Physics, University of Tartu, Riia 142, Tartu 51014, Estonia \\ \\
 Margus Saal\thanks{margus@fi.tartu.ee} \\
 Tartu Observatory, T\~oravere 61602, Estonia}
\date{}
\maketitle

\begin{abstract}
We consider the potential dominated era of Friedmann-Lema\^{\i}tre-Robertson-Walker flat
cosmological models in the framework of general Jordan frame scalar-tensor theories of gravity 
with arbitrary coupling functions,
and focus upon the phase space of the scalar field. 
To study the regime suggested by the local weak field tests (i.e. close to the so-called limit of general relativity)
we propose a nonlinear approximation scheme, 
solve for the phase trajectories, and provide a complete classification of possible phase portraits.
We argue that the topology of trajectories in the nonlinear approximation is representative of those
of the full system, and thus can tell for which scalar-tensor models general relativity functions as an
attractor.
\end{abstract}



\vspace{1cm}

\section{Introduction}

The unknown source of observed present day acceleration 
of the Universe, called dark 
energy, is inspiring thorough investigations of different extensions of general relativity (GR)
and $\Lambda$CDM cosmology (for recent reviews see Refs. \cite{de:theory}). 
The scalar-tensor theory of gravity (STG) \cite{books} offers one such consistent possibility. 
Besides the usual spacetime metric tensor $g_{\mu\nu}$ it employs a scalar field $\Psi$, playing the role of a variable 
gravitational ``constant", to describe the 
gravitational interaction.
In the Jordan frame STG is specified by two functions \cite{flanagan}, 
e.g. a coupling $\omega(\Psi)$ and a scalar potential $V(\Psi)$.
In fact, a wide class of theories of gravitation,
including higher order theories \cite{sotiriou}, theories of variable speed of light \cite{magueijo}, 
as well as low energy approximations of brane world models and string theories \cite{gasperini} can be cast into the general form of STG.

The weak field tests \cite{ppn} pose a restriction to all 
alternative models of gravity including STG, since the Universe around us tends to be 
described by the Einstein tensorial gravity very precisely \cite{constraints}.
This means that only those STG models are physically 
viable which in their late time cosmological evolution
imply local consequences very close to those of GR.
Several authors have studied how general relativity acts as an attractor for a wide class of STGs, 
such that the Solar System weak field (PPN) constraints spontaneously come to be satisfied at late times 
\cite{dn,bp,asymtotics}.
In our recent papers \cite{meie4, meie5} we have proposed a limiting process for the scalar 
field which describes scalar-tensor cosmological models relaxing to satisfy the Solar System 
constraints, with an indication for which classes of STGs the attractor behavior is realized. 

The methods of dynamical systems have proved to be a useful tool when explicit analytic solutions are hard to find. 
In STG cosmology several authors have performed the analysis for different specific 
choices of the coupling and potential \cite{stg:dynsys}, while Refs. \cite{faraoni1,burd_coley,faraoni3,meie4} study the phase space and dynamics in the general case.

The STG phase space point corresponding to the limit of GR is peculiar 
in the sense that the standard linearization process there is hampered by ratios 
which turn out to be indeterminate.
In our previous studies \cite{meie4, meie5} we assumed that these indeterminate terms
vanish, which allows to treat this singular point as a standard fixed point.
In the present paper we focus upon the potential dominated era of 
Friedmann-Lema\^{\i}tre-Robertson-Walker (FLRW) flat cosmological models
in the phase space of the (decoupled) scalar field $\Psi$  and its (cosmological) time derivative ${\dot \Psi} \equiv \Pi$,
and propose an approximation which takes into account all possible 
finite values of these indeterminate ratios, thus preserving 
the leading nonlinear term in the field equations.
We give a comprehensive description of the phase space trajectories of the approximate nonlinear system near the GR point and classify all trajectories allowed by the parameters of the theory.
While the topology of trajectories differs in linear and nonlinear approximation, there is a correspondence in the final asymptotics, i.e. whether the trajectories end up at the GR point or are repelled from it. 
We argue that the nonlinear system accurately captures the key properties of the full system of STG equations near this singular point, and the topology of trajectories of the nonlinear approximation 
is representative of those of the full system.
Therefore for any given STG model with a reasonable coupling $\omega(\Psi)$ and scalar potential $V(\Psi)$ our results predict whether GR is an attractor.

The paper is organized as follows.
In the next section we review very briefly the scalar-tensor theory of gravity
in the potential dominated era,
write down the field equations of FLRW cosmology 
in the form of a dynamical system, and make some general remarks about the phase space
including the singular GR point.
In Sect. \ref{ae} we introduce the approximation method 
and present linear and nonlinear systems of equations 
in the neighbourhood of the singular point.
In Section \ref{pt} 
we present solutions (phase trajectories) of the nonlinear system and 
a classification of trajectories,
summarized in Table 1 and 
illustrated on Figure 1.
In Section \ref{trivial} 
our claims are backed up by a simple example.
Finally, Section \ref{disc} provides a summary and some remarks for future work. 


\section{Full equations and the phase space} \label{fe}

We consider  a general scalar-tensor theory
in the Jordan frame given by the action functional
    \beq \label{jf4da}
S  = \frac{1}{2 \kappa^2} \int d^4 x \sqrt{-g}
        	        \left[ \Psi R(g) - \frac{\omega (\Psi ) }{\Psi}
        		\nabla^{\rho}\Psi \nabla_{\rho}\Psi
                  - 2 \kappa^2 V(\Psi)  \right]\,.
\eeq 
(We have not included 
the matter contribution to the action, 
i.e. we consider only the potential dominated epoch.)
Here $\omega(\Psi)$ is a coupling function 
(we assume $2 \omega (\Psi) + 3 \geq 0 $ to avoid ghosts in the
Einstein frame, see  e.g.  Ref. \cite{polarski}), 
$V(\Psi)\geq 0$ is a scalar potential, $\nabla_{\mu}$ 
denotes the covariant derivative with respect to the metric 
$g_{\mu\nu}$, and $\kappa^2$ is the non-variable part of
the effective gravitational constant $\frac{\kappa^2}{\Psi}$. 
In order to keep it positive
we assume that $0 < \Psi < \infty$.

The field equations for the flat Friedmann-Lema\^{\i}tre-Robertson-Walker
(FLRW) line element 
\be
ds^2=-dt^2 + a(t)^2 \left(dr^2 + r^2 (d\theta ^2 + \sin ^2 \theta d\varphi ^2)\right)
\ee
read 
\beq 
\label{00}
H^2 &=& 
- H \frac{\dot \Psi}{\Psi} 
+ \frac{1}{6} \frac{\dot \Psi^2}{\Psi^2} \ \omega(\Psi) 
+ \frac{\kappa^2}{3} \frac{V(\Psi)}{\Psi} \,, 
\\ \nonumber \\
\label{mn}
2 \dot{ H} + 3 H^2 &=& 
- 2 H \frac{\dot{\Psi}}{\Psi} 
- \frac{1}{2} \frac{\dot{\Psi}^2}{\Psi^2} \ \omega(\Psi) 
- \frac{\ddot{\Psi}}{\Psi} 
+ \frac{\kappa^2}{\Psi} \ V(\Psi) \,, 
\\ \nonumber \\
\label{deq}
\ddot \Psi &= & 
- 3H \dot \Psi 
- \frac{1}{2\omega(\Psi) + 3} \ \frac{d \omega(\Psi)}{d \Psi} \  \dot {\Psi}^2 
+ \frac{2 \kappa^2}{2 \omega(\Psi) + 3} \ \left[ 2V(\Psi) - \Psi \ 
\frac{d V(\Psi)}{d \Psi}\right] ,
\eeq 
where $H \equiv \dot{a} / a$.

By  defining $ \dot{\Psi} \equiv \Pi$, the
equations  (\ref{00})--(\ref{deq}) can be considered as an autonomous dynamical system,  
which is characterized by three variables $( \Psi, \Pi, H )$. However,
one of them is algebraically related to the others via the Friedmann equation (\ref{00}). We choose to eliminate the
Hubble parameter $H$ and thus reduce the system to two dimensions $( \Psi, \Pi)$.
Upon  introducing the functions 
\be  \label{A_W}
A(\Psi) \equiv \frac{d}{d\Psi} \left(\frac{1}{2\omega(\Psi)+3} \right)\,, \qquad
W(\Psi) \equiv 2 \kappa^2  \left(2V(\Psi)-\frac{dV(\Psi)}{d\Psi}\Psi \right)\,
\ee
the dynamical system reads
\beq
\label{dynsys_x}
\dot{\Psi} &=& \Pi \,, \\ 
\label{dynsys_y} 
\dot{\Pi} &=& \left( \frac{3}{2\Psi} + \frac{1}{2} A(\Psi) (2\omega(\Psi)+3)\right) \Pi^2 \\ \nonumber
&& \pm \frac{1}{2\Psi} \sqrt{3( 2\omega(\Psi) + 3) \Pi^2   
+ 12 \kappa^2 \Psi V(\Psi)} \ \Pi   
 +\frac{W(\Psi)}{2\omega(\Psi)+3}  \,.
\eeq 
Its regular phase trajectories and fixed points have been considered in Refs. \cite{meie4, faraoni3}. 

Even if the scalar field rules the cosmological evolution,
in the context of Solar System experiments we can reasonably assume that the energy density of the 
potential is negligible
in comparison with the local matter density. Then the standard PPN analysis
gives a condition for the present cosmological background value of the scalar field \cite{ppn},  
(a) $\frac{1}{2\omega(\Psi)+3} \rightarrow 0$. 
It is also evident that the effective gravitational constant is virtually immutable in time \cite{bp}, which provides the second condition,
(b) ${\dot \Psi} \equiv \Pi \rightarrow 0$. 
Let us denote $ \Psi_{\star} $ the value of the scalar field  where the coupling function $\omega(\Psi)$ has a singular ``peak'', or more precisely $\frac{1}{2\omega(\Psi_{\star})+3} = 0$. 
Also, let $\Pi_{\star}$ be its vanishing time derivative, $ \Pi_{\star} = 0$.
Then we may cautiously call the phase space point ($ \Psi = \Psi_{\star}$, $ \Pi = \Pi_{\star}$) a `GR point', since the STG solutions which can pass the local weak field tests and behave close enough to the ones of general relativity in terms of local observations, 
 necessarily lie in the vicinity of this point. 
 
The value $\Psi_{\star}$ poses a caveat, though.
On the right hand side of Eq. (\ref{dynsys_y})
the $(2\omega(\Psi_{\star})+3) \Pi^2$ terms 
are diverging if $\Pi \not= 0$ and indeterminate if $\Pi = \Pi_{\star} =0$.
This means that the whole set $\Psi =\Psi_{\star}$ at arbitrary $\Pi$ is excluded from the (open) domain of definition 
of Eq. (\ref{dynsys_y}). However, in what follows we find 
solutions of general approximated equations (Sect. \ref{pt}) and of a specific full equation 
(Sect. \ref{trivial}) for phase trajectories which 
smoothly reach or pass through the point ($\Psi_{\star}$, $\Pi_{\star}$).  
Thus we are justified to add the GR point to the open domain of definition as a boundary point.  
(Note that analogous conclusion can be inferred from our analysis of regions in the phase space 
accessible to phase trajectories in a general STG \cite{meie4}.) 

The phase portrait of the dynamical system (\ref{dynsys_x}), (\ref{dynsys_y})
is drawn by the solutions (trajectories) of 
\beq  \label{orbit_ex}
\frac{d\Pi}{d\Psi} &=& \left( \frac{3}{2\Psi} + \frac{1}{2} A(\Psi) (2\omega(\Psi)+3)\right) \Pi 
   \\ \nonumber
 &&\pm  \frac{1}{2\Psi} \sqrt{3( 2\omega(\Psi) + 3) \Pi^2   
+ 12 \kappa^2 \Psi V(\Psi)}     +\frac{W(\Psi)}{(2\omega(\Psi)+3) \Pi}  \,,
\eeq   
while the direction of the flow along them is determined by Eq. (\ref{dynsys_x}) as usual
(towards increasing $\Psi$ for $\Pi>0$ and towards decreasing $\Psi$ for $\Pi<0$).
Realizing that $\frac{d\Pi}{d\Psi}$ gives the slope of the tangent to a phase trajectory,
a few characteristic features of the phase portrait outside the singular point ($\Psi_{\star}$, $\Pi_{\star}$) can be immediately inferred.

First, on the horizontal line $\Pi =0$ ($\Psi$-axis) the tangents of trajectories are vertically aligned if $W(\Psi) \neq 0$, 
since $\frac{d\Pi}{d\Psi}$ diverges due to the last term in Eq. (\ref{orbit_ex}).
The inverted derivative $\frac{d\Psi}{d\Pi}|_{\Pi=0}$ vanishes, while the sign of $\frac{d^2\Psi}{d\Pi^2}|_{\Pi=0} \sim \frac{1}{W(\Psi)}$ indicates the direction of the flow along the trajectories:
passing from $\Pi>0$ to $\Pi<0$ if $\frac{d^2\Psi}{d\Pi^2}|_{\Pi=0} < 0$ and vice versa if $\frac{d^2\Psi}{d\Pi^2}|_{\Pi=0} > 0$. 
The case of quadratic potential is special, as now $W(\Psi) \equiv 0$, and Eqs. (\ref{dynsys_x}), (\ref{dynsys_y}) reveal 
that besides the singular (indeterminate) point ($\Psi_{\star}$,$\Pi_{\star}$),  
all points on the line $\Pi =0$ are fixed points, i.e. the trajectories do not pass through, but 
either begin or end there.
   
Second, approaching on the vertical line $\Psi =\Psi_{\star}$ the tangents
of phase trajectories turn again vertical due 
to $(2\omega(\Psi)+3)$ blowing up as $\Psi \rightarrow \Psi_{\star}$. 
As has been argued above,
the line $\Psi =\Psi_{\star}$ ($\Pi \neq 0$) itself does not belong to the domain of the definition of the system, and 
here we see that the trajectories acknowledge this fact 
by not running into that line, but radically deflecting ``up'' or ``down'' instead,
depending on the sign of 
$\frac{d\Pi}{d\Psi}|_{\Psi \rightarrow \Psi_{\star}}$.

These two features qualitatively control the behavior of trajectories around the GR point.
Consider for instance the region $\Psi<\Psi_{\star}$, $\Pi>0$ where the flow is directed towards
$\Psi_{\star}$. First, if $\frac{d^2\Psi}{d\Pi^2}|_{\Pi=0} < 0$ at least some of these trajectories will cross over to the $\Pi<0$ belt and then flow away from $\Psi_{\star}$,
effectively displaying a ``saddle'' type of behavior. 
On the other hand, if $\frac{d^2\Psi}{d\Pi^2}|_{\Pi=0} > 0$ there is no other option 
for the trajectories
but to persist in flowing towards $\Psi_{\star}$, while additional trajectories coming
over from the $\Pi<0$ region join their course. Second, near the $\Psi=\Psi_{\star}$ line
if $\frac{d\Pi}{d\Psi}|_{\Psi \rightarrow \Psi_{\star}}$ is positive, the flow is pushed 
``upwards'' to $\Pi \rightarrow \infty$, while
if $\frac{d\Pi}{d\Psi}|_{\Psi \rightarrow \Psi_{\star}}$ is negative, the push is ``downwards'' 
towards smaller values of $\Pi$. In the latter case the outcome again depends on how the flow is 
directed on the $\Pi=0$ line, viz. if $\frac{d^2\Psi}{d\Pi^2}|_{\Pi=0} < 0$ the trajectories 
can cross over to the $\Pi<0$ region at any $\Psi \neq \Psi_{\star}$ and flow away, 
while if $\frac{d^2\Psi}{d\Pi^2}|_{\Pi=0} > 0$
the trajectories would have nowhere else to go but to hit the 
point ($\Psi_{\star}$, $\Pi_{\star}$).
A similar reasoning can be put forth for $\Psi>\Psi_{\star}$ as well.
It may also happen that as the potential $V(\Psi)$ varies,
the quantity $\frac{d^2\Psi}{d\Pi^2}|_{\Pi=0} \sim \frac{1}{W(\Psi)}$ may be positive or negative 
depending on the value of $\Psi$ 
and the global picture gets rather complicated.

What can be said about the trajectories at the singular (indeterminate) point
($\Psi_{\star}$, $\Pi_{\star}$)?
The logic of the phase space 
tells that passing through this point is possible
from a region of the phase space where the flow is directed towards the point 
into a region of the phase space where the flow is directed away from the point.
Under a reasonable assumption that (at least) the (physically relevant) 
solutions are continuous and smooth there are two such possibilities. 
First, the solutions may ``slip through'' $\Psi_{\star}$ 
from the region $\Psi<\Psi_{\star}$, $\Pi>0$ 
($\Psi>\Psi_{\star}$, $\Pi<0$) to the region $\Psi>\Psi_{\star}$, $\Pi>0$ 
($\Psi<\Psi_{\star}$, $\Pi<0$) 
if $\frac{d\Pi}{d\Psi}|_{\Psi \rightarrow \Psi_{\star}, \Pi \rightarrow \Pi_{\star}} = 0$,
i.e. the tangent of the trajectory is aligned horizontally.
Second, the solutions may ``bounce back'' from $\Psi_{\star}$ 
from the region $\Psi<\Psi_{\star}$, $\Pi>0$ 
($\Psi>\Psi_{\star}$, $\Pi<0$) to the region $\Psi<\Psi_{\star}$, $\Pi<0$ 
($\Psi>\Psi_{\star}$, $\Pi>0$) 
if $\frac{d\Psi}{d\Pi}|_{\Psi \rightarrow \Psi_{\star}, \Pi \rightarrow \Pi_{\star}} = 0$,
i.e. the tangent of the trajectory is aligned vertically.
Any trajectory hitting this point under a tangent
which is neither horizontal nor vertical
can not pass through the point, but must terminate there.

\section{Approximate equations} \label{ae}

The equations (\ref{dynsys_x}), (\ref{dynsys_y}) cannot be integrated 
without specifying the two
arbitrary functions $\omega(\Psi)$ and $V(\Psi)$.
But being interested in the behavior of solutions close to the GR point 
($\Psi_{\star}$, $\Pi_{\star} $) we can still proceed
by considerning an approximation which maintains the key properties of the full system
near this point.
Although the full equations become singular (indeterminate) at ($\Psi_{\star}$, $\Pi_{\star} $), 
we assume that 
the Taylor expansions of the functions $\omega (\Psi)$, $V(\Psi)$ are possible there.    

Let us focus around the GR point,
\be
\label{expans}
\Psi = \Psi_{\star} + x \,, \qquad \Pi = \Pi_{\star} + y = y \,, 
\ee
where $x$ and $y$ span the neighbourhood of first order small distance
from ($\Psi_{\star}$, $\Pi_{\star} $). 
As phase space variables $x$ and $y$ are independent from each other,
and so their ratio $y/x$ is indeterminate at ($x = 0$, $y= 0$).
The meaning of this indeterminacy is perhaps better
illuminated in the polar coordinates ($\rho, \theta $), where
the radius $\rho$ is a first order small quantity, but 
$y/x \equiv \tan \theta \in (-\infty, \infty)$  
becomes infinitely multivalued at the origin. 

We can Taylor expand 
\be
\frac{1}{2\omega(\Psi)+3} = \frac{1}{2\omega(\Psi_{\star})+3}+ A_{\star} x+ ...\approx A_{\star} x \,,     
\ee
and
\be \label{y2/x} 
(2\omega(\Psi)+3) \Pi^2 = \frac{y^2}{0 + A_{\star} x + ...} = \frac{y^2}{A_{\star} x}\left(1 + O(x)\right) 
\approx \frac{y^2}{A_{\star} x} \,,  
\ee
where $A_{\star} \equiv A(\Psi_{\star})$.
In order to keep the expansion under better control we have introduced here two additional conditions: 
(c)$A_{\star} \not= 0$, and (d) $\frac{1}{2\omega(\Psi)+3} $ is differentiable at  $\Psi_{\star}$
($A_{\star}$ and higher derivatives do not diverge) \cite{meie4, meie5}. 
Although these assumptions somewhat constrain the possible forms of $\omega$,
we are still dealing with a wide and relevant class of theories.
In fact, the set (a)-(d) guarantees, that the second condition for 
the cosmological background value of the scalar field
arising in the PPN analysis of STG \cite{ppn}, 
$\frac{1}{(2\omega(\Psi)+3)^3}\frac{d\omega}{d\Psi} \rightarrow 0$, 
is automatically satisfied \cite{meie4}.

To simplify the notation, 
let us denote the values of some functions at  ($\Psi_{\star}$, $\Pi_{\star}$) as
\be
\label{def}
 C_1 \equiv \pm \sqrt{\frac{3 \kappa^2 V(\Psi_{\star})}{\Psi_{\star}}} \,, \qquad
C_2 \equiv A_{\star}\, W_{\star} \,, 
\ee
where $W_{\star} \equiv W(\Psi_{\star})$ and $V(\Psi_{\star})\geq 0$. 
The three constants $A_{\star}$, $W_{\star}$, $C_1$ determine the leading terms in expansions of 
the two functions $\omega(\Psi)$, $V(\Psi)$ which specify a STG.
Now the expansion of the solution for $H$ of the Friedmann constraint (\ref{00}) reads
\beq  \label{H_approx}
H &=& -\frac{\Pi}{2\Psi} \pm \sqrt{(2\omega(\Psi)+3)\frac{\Pi^2}{12 \Psi^2} + \frac{\kappa^2 V(\Psi)}{3 \Psi}} \\ \nonumber
&\approx& - \frac{y}{2\Psi_{\star}} + 
\frac{C_1}{3}\left[1 + \frac{3 \kappa^2}{2C_1^2} \frac{d}{d \Psi} \left( \frac{V}{\Psi}\right)_{\star} x + 
\frac{3}{8 C_1^2 \Psi_{\star}A_{\star}} \ \frac{y^2}{x} + ... \right]\,.
\eeq
This explains the introduction of the $\pm$ sign in the definition of $C_1$ in Eq. (\ref{def}),
as near the GR point 
($x=0, y=0$) a positive constant, $C_1 > 0$, describes an expanding de Sitter Universe, while a negative one, 
$C_1 < 0$, describes a contracting de Sitter Universe.



Having outlined the method of approximation 
in the neighbourhood of 
($\Psi_{\star}$, $\Pi_{\star}$),
let us apply it for the system (\ref{dynsys_x}), (\ref{dynsys_y}). 
If we assume, motivated by the condition (b), that physically relevant trajectories 
linger in the region close to the $x$-axis, i.e 
$\frac{y}{x} = \tan \theta $ being first order small, 
then in the expansion of Eq. (\ref{dynsys_y}) 
only the terms linear in $x$ and $y$ survive at the first order,
while terms like $\frac{y^2}{x}$ (cf. Eq. (\ref{y2/x})) can be dropped. 
This was the assumption implicit in our earlier analysis \cite{meie4, meie5}.
In this case, denoting the variables ${\tilde x}$ and ${\tilde y}$, the 
approximation of (\ref{dynsys_x}), (\ref{dynsys_y}) yields a linear system
\beq    
\label{lin_x}
\dot{{\tilde x}} &=& {\tilde y}  \,, \\ 
\label{lin_y} 
\dot{{\tilde y}} &=& C_2 \ {\tilde x} - C_1 \ {\tilde y} \,,
\eeq
which, of course, is equivalent to a general second order linear homogeneous differential equation
\be
\label{lin_eq}
{\ddot {\tilde x}} + C_1 \ {\dot {\tilde x}} - C_2 \ {\tilde x} = 0 \,.
\ee
Its phase space analysis is well known, there is a fixed point at ($x=0, y=0$)
whose type depends on the values 
of the constants $C_1$ and $C_2$ (see e.g \cite{arnold_2}).

However, in a more general case
we must 
recognize the term $\frac{y^2}{x}$ as being the same order as $x$ and $y$. 
In other words,
we consider all finite values of $\tan \theta$, and exclude only its infinite value 
on the $y$-axis which is outside the domain of definition of the system as said before.
Thus, keeping the term $\frac{y^2}{x}$ in the approximation of (\ref{dynsys_x}), (\ref{dynsys_y}),
we obtain a nonlinear system
\beq    
\label{mlin_x}
\dot{x} &=& y \,, \\ 
\label{mlin_y} 
\dot{y} &=&  \frac{y^2}{2x} - C_1 \ y +C_2 \ x \,.
\eeq
The corresponding second order nonlinear differential equation reads
\be
\label{mlin_eq}
{\ddot x} + C_1 \ {\dot x} - C_2 \ x =  \frac{{\dot x}^2}{2x} \,.
\ee
Note that as distinct from eqs. (\ref{lin_x}), (\ref{lin_y}), 
in this case  point  ($x=0$, $y=0$) is not a fixed point with $\dot{x} = 0$, $\dot{y} =0$
any more,  
but it is a singular point with an indeterminate and possibly multivalued term $\frac{y^2}{2x}$ 
in eq. (\ref{mlin_y}).


\section{Phase trajectories } \label{pt}

The phase trajectories for the nonlinear approximate system (\ref{mlin_x}), (\ref{mlin_y}) 
are determined by the equation
\be \label{eq_orb_ml}
\frac{dy}{dx} = \frac{y}{2x} - C_1 + \frac{x}{y} \ C_2 \,.
\ee
Its solutions 
\be \label{orbit_ml}
|x| K = \left|\frac{1}{2} y^2 + C_1 yx - C_2 x^2\right| \exp(- C_1 f(u)) \,,  \quad  u \equiv \frac{y}{x} \,,
\ee
depend on the sign of the expression $C_1^2 + 2C_2 \equiv C$, as the function $f(u)$ is given by 
\beq  \label{fu}
f(u) &=& \frac{1}{\sqrt{C}} \ln  \left|\frac{u + C_1 - \sqrt{C}}{u + C_1 + \sqrt{C}} \right| \qquad  \qquad {\rm if} \quad C > 0 \,,   \nonumber \\      
     &=& - \frac{2}{u + C_1}  \qquad \qquad \qquad \qquad \qquad  {\rm if} \quad  C = 0 \,, \nonumber \\
     &=&  \frac{2}{\sqrt{|C|}} \left( \arctan \frac{u + C_1}{\sqrt{|C|}} + n \pi \right) \, \quad   {\rm if} \quad C < 0 \,.  
\eeq
Here $K$ is a constant of integration which identifies the trajectory according to initial data ($x_0$, $y_0$).
Note that if we choose initial conditions from the allowed region ($\tan \theta$ is finite), 
then our premise $\frac{y^2}{x} \sim y$ at deriving 
approximate equations (\ref{mlin_x}), (\ref{mlin_y}) is always valid, i.e. we get a small constant of integration, 
$K < 1$. 
 
In general, the right hand side of Eq. (\ref{eq_orb_ml}) can be written 
as a quotient of two second order homogeneous polynomials; 
a qualitative classification of the solutions of differential equations 
of this type was given by Lyagina \cite{umn} long time ago. 
In a nutshell, the phase portraits for different values of the constants $C_1$ and $C_2$
classify according to the number of sectors which form on the phase space 
around the origin ($x=0, y=0$),
and the topology of trajectories which inhabit these sectors.
The sectors are separated by the boundary $x=0$ and invariant directions.
The latter are lines $y=k x$ 
where the constant $k$ is a real solution of an algebraic equation
\be \label{invdir_eq}
k= \frac{k}{2} -C_1 + \frac{C_2}{k} \,,
\ee
i.e straight trajectories $y = (- C_1 \pm \sqrt{C}) x$ satisfying (\ref{eq_orb_ml}).
All possible options are listed in Table~1 and graphically depicted on Figure~1.

If $C>0$ and $C_2 \not= 0$ three directions divide the phase space into
six topologically distinct sectors. The sectors can be 
elliptic where all trajectories start from the origin and get back to the origin,
hyperbolic where all trajectories flow towards the origin 
but turn back before reaching it,
or parabolic where all trajectories either start from afar
and flow to the origin (stable case) or start from the origin and flow away (unstable case).

If $C_1 \not= 0$ and $C_2 = 0$, i.e. the potential has a special form $V(\Psi_{\star}) \not= 0$,
$\left(2V(\Psi_{\star})-\frac{dV(\Psi)}{d\Psi}|_{\Psi_{\star}} \Psi_{\star} \right)= 0$, then 
it follows from the original dynamical system (\ref{mlin_x}), (\ref{mlin_y}) that 
the entire $x$-axis (the point $x=0$ excluded) is populated by degenerate fixed points.
There are four sectors. The two sectors which contain the $x$-axis are special,
and do not properly belong to neither elliptic, hyperbolic or parabolic class, as the flow 
there is dominated by the cohort of fixed points lying on the $x$-axis.
Let us provisionally call these `sectors of degenerate fixed points'.

If $C=0$ there are four sectors. If both $C_1 = 0$ and $C_2= 0$, the $x$-axis consists again
of fixed points, while the generic trajectories are parabolas $2 K |x| = y^2$.
If $C < 0$ there are no real solutions to Eq. (\ref{invdir_eq}) and all we get are two elliptic sectors
on both sides of the $y$-axis. 

It is worth pointing out here, that the phase portraits of the nonlinear approximation
display the same basic characteristic features we inferred about the solutions of the 
full system (\ref{orbit_ex}) before. First, on the horizontal axis ($y=0$) the
tangents of the trajectories are vertically aligned if $C_2 \neq 0$, and the 
direction of the flow across $y=0$ is determined by the sign of $\frac{d^2 x}{d y^2}|_{y=0}$.
If $C_2 = 0$ the horizontal axis is populated by fixed points. 
Second, next to the vertical axis ($x=0$) the trajectories turn vertical and 
do not cross or intersect with the $x=0$, $y \neq 0$ line,
deemed to be outside of the domain of definition of the system.

What happens at the origin ($x=0, y=0$) where the sectors meet, needs an extra consideration.
Inspection of the phase portraits on Figure~1 shows that
in all cases there are multiple trajectories (identified by different values of $K$)
which all reach the point in question.
Although our solutions for the trajectories, 
Eq. (\ref{orbit_ml}), are given in terms of the phase space variables only and
do not include time as an explicit parameter, 
the considerations presented in the end of Sect.~2 allow to draw
some qualitative conclusions.
First, there was a logical possibility of trajectories ``slipping through'' the origin
so that $x$ changes its sign along a trajectory. 
It is evident from the phase portraits that this option is not
realized in any of the cases, as none of the trajectories
has $\tan \theta =0$ at this point.
On the other hand, the second possibility, where the trajectories could 
``bounce back'' from the origin so that $y$ changes its sign along a trajectory,
is common to all cases, for there is always a class of trajectories
whose tangent is vertically aligned at this point. 
Despite the fact that there seems to be loss of 
predictability here 
(the initial condition $x_0=0$, $y_0=0$ does not fix the constant $K$ uniquely),
it would be natural to continue all such trajectories through this point
keeping the same $K$ along them. 
Finally, those trajectories which reach the origin under finite $\tan \theta$
must either begin or end their flow at this point, 
like it happens at a regular fixed point.

It is also instructive to compare the 
phase portraits in the 
linear and nonlinear approximations. 
In the linear approximation (\ref{lin_x}), (\ref{lin_y}) the phase diagram is 
determined by 
\be \label{eq_lin}
\frac{d{\tilde y}}{d{\tilde x}} =  - C_1 + \frac{{\tilde x}}{{\tilde y}} \ C_2 \,.
\ee
Its solutions are
\be  \label{orbit_lin}
{\tilde K} = \big| {\tilde y}^2 + C_1 {\tilde x}{\tilde y}  - C_2 {\tilde x}^2 \big| \exp(- \frac{C_1}{2} f(\tilde u)) \,, \qquad 
{\tilde u} \equiv \frac{2{\tilde y}}{{\tilde x}} 
\ee
where $f({\tilde u})$ is given by the same expression (\ref{fu}),
while instead of $C$ 
the solution is set by the sign of the expression ${\tilde C} \equiv C_1^2 + 4C_2 $. 
Now the invariant sectors are separated by the lines $\tilde{y}=\tilde{k} \tilde{x}$, where
the constant $\tilde{k}$ is a real solution of 
\be \label{invdir_eq}
\tilde{k}= -C_1 + \frac{C_2}{\tilde{k}} \,,
\ee
i.e. $\tilde{k}=\frac{1}{2}(-C_1 \pm \sqrt{\tilde{C}})$.
There is no indeterminacy in Eqs. (\ref{lin_x}), (\ref{lin_y}) and 
the point 
(${\tilde x} =0$, ${\tilde y} =0$) figures as a regular fixed point.
The corresponding phase portraits in the neighbourhood of 
the origin
can be classified according to a standard analysis \cite{meie4, meie5}, 
the results being presented in Table 1.

The phase portraits in the nonlinear and linear approximation look markedly different,
as in the linear case the line ${\tilde x}=0$ is not a boundary, and 
the trajectories for nodes, focuses etc.
are not hindered from crossing it.
Yet, modulo a factor of 2 in $C$ vs 4 in $\tilde{C}$, 
there is an overall correspondence in the classification,
i.e. each distinct class in the linear case is matched with a 
distinct class in the nonlinear case (cf. Table 1). 

This correspondence in the classification occurs due to an (accidental)
property that the nonlinear system (\ref{mlin_x}), (\ref{mlin_y}) for ($x, y$) 
can be formally 
obtained from the linear system (\ref{lin_x}), (\ref{lin_y}) for (${\tilde x}, {\tilde y}$) 
by a replacement $C_2 \rightarrow  \frac{C_2}{2}$ 
(to get $\tilde{C} \rightarrow C$) and a simple transformation    
${\tilde x} = \sqrt{|x|} $, ${\tilde y} = \frac{y}{2 \sqrt{|x|}}$.
The latter can be understood as a mapping between the corresponding phase spaces. 
In particular, it squeezes the whole ${\tilde y}$-axis 
to a single point ($x=0$, $y = 0 $), 
which manifests as the indeterminacy of the nonlinear system at
this point.
Another characteristic property 
of the map is the fact that the whole phase space (${\tilde x}, {\tilde y}$) is mapped only 
on one half of the phase space ($x, y$), but covering it 
twice. Since the transformation contains absolute value $|x|$, there are two separate images symmetric with respect to the $y$-axis.     

Although the mapping between the linear and nonlinear systems seems to be
only a mathematical coincidence and not a consequence of the fact that both 
the linear and nonlinear systems originate as approximations to the
full system of equations
(the nonlinear being a more general and refined one in this respect),
we can nevertheless utilize this correspondece to
unravel some useful information.
Namely, the trajectories which cross the $\tilde{x}=0$ line 
at arbitrary $\tilde{y}$ in the linear case
are mapped onto the trajectories which vertically hit the point ($x=0, y=0$), 
and thus, as the former pass through the $\tilde{x}=0$ line 
the mapping suggests that also the latter must pass through the point ($x=0, y=0$),
hence supporting our reckoning above. 
Moreover, for $C<0$ or $\tilde{C}<0$ the trajectories 
depend not only on the real constant $K$ or $\tilde{K}$, but also on the integer $n$ 
due to the periodicity of $\arctan$ in Eq (\ref{fu}). 
Here in the linear case we have either a
stable focus (spiralling trajectories flowing inwards into the origin)
or unstable focus (spiralling trajectories flowing outwards from the origin),
while $n$ decreases by 1 on each occasion 
when a trajectory flows through the $\tilde{y}$ axis.
The mapping tells now, that also in the nonlinear case
$n$ must decrease by 1 on each occasion 
when a trajectory flows through the point ($x=0, y=0$), 
thus the overall picture being trajectories looping closer and closer or
farther and farther from the origin. (On Figure 1 the diagrams 3.a and 3.c 
depict a trajectory with some fixed $K$ but different values of $n$,
the direction of the flow is indicated by the increasing number of arrows on 
a trajectory loop.)

Thus (the mismatch of factor 2 vs 4 notwithstanding) 
besides the correspondence in the classification of the phase portraits,
the linear and nonlinear approximations also share a qualitative correspondence in
final asymptotic state of the flow, i.e. whether it ends up at 
the origin, or departs away from it. 
To summarize the results, it turns out that the GR point is an attractor for 
the asymptotic flow of all trajectories only if $C_1 >0$ and $C_2<0$ 
(cases 1.c, 2.a, 3.a). 
If $C_1 >0$ and $C_2=0$ all trajectories flow to the line $\Psi \neq \Psi_{\star}$,
$\Pi=0$ instead (case 1.b). 
If $C_1 =0$ and $C_2<0$ all trajectories loop through the GR point
oscillating back and forth (nonlinear case 3.b), 
or if $C_1 <0$ and $C_2<-\frac{C_1}{2}$ they oscillate further and further
(nonlinear case 3.c). 
For the rest of the values of $C_1$ and $C_2$ all trajectories eventually flow
away from the GR point.


\begin{table}
\begin{flushleft}
\begin{tabular}{llllll} 
No.  &Parameters                 &Fixed point,                               &Topology of trajectories, \\
     &LS:$N=4$;\ NLS:$N=2$       &linear system                              &nonlinear system   
\vspace{2mm}\\
\hline \hline \\
1. & $C_{1}^2 + N C_{2} > 0$
\vspace{2mm}\\
\cline{1-2} 
\vspace{2mm}\\
1.a & $C_{1} > 0$ \qquad \quad \ \ $C_{2} > 0$           &saddle           &2 hyperb., 2 st. \& 2 unst. parab. sectors \vspace{2mm}\\ 
1.b & $C_{1} > 0$ \qquad \quad \ \ $C_{2} = 0$           &non-hyperbolic   &1 stable \& 1 unstable parabolic sector,               \vspace{2mm}\\
 & & &2 stable sectors of degenerate fixed points               \vspace{2mm}\\
1.c & $C_{1} > 0$ \  $-\frac{C_{1}^2}{N}< C_{2} < 0$     &stable node      &2 elliptic, 4 stable parabolic sectors              \vspace{2mm}\\
1.d & $C_{1} = 0$ \qquad \quad \ \  $C_{2} > 0$          &saddle           &2 hyperb., 2 st. \& 2 unst. parab. sectors \vspace{2mm}\\
1.e & $C_{1} < 0$ \qquad \quad \ \ $C_{2} > 0$           &saddle           &2 hyperb., 2 st. \& 2 unst. parab. sectors \vspace{2mm}\\
1.f & $C_{1} < 0$ \qquad \quad \ \ $C_{2} = 0$           &non-hyperbolic   &1 stable \& 1 unstable parabolic sector,                \vspace{2mm}\\
 & & &2 unst. sectors of degenerate fixed points               \vspace{2mm}\\
1.g & $C_{1} < 0$ \ $-\frac{C_{1}^2}{N}< C_{2} < 0$      &unstable node    &2 elliptic, 4 unstable parabolic sectors            \vspace{2mm}\\
\hline \\
2. & $C_{1}^2 + N C_{2} = 0$
\vspace{2mm}\\
\cline{1-2} 
\vspace{2mm}\\
2.a & $C_{1} > 0$  \qquad  $C_{2} = -\frac{C_{1}^2}{N}$  &stable node      &2 elliptic, 2 stable parabolic sectors    \vspace{2mm}\\ 
2.b & $C_{1} = 0$  \qquad  $C_{2} = 0 $                  &free motion      &2 stable \& 2 unstable parabolic sectors     \vspace{2mm}\\
2.c & $C_{1} < 0$  \qquad  $C_{2} = -\frac{C_{1}^2}{N}$  &unstable node    &2 elliptic, 2 unstable parabolic sectors  \vspace{2mm}\\
\hline \\
3. & $C_{1}^2 + N C_{2} < 0$
\vspace{2mm}\\
\cline{1-2} 
\vspace{2mm}\\
3.a & $C_{1} > 0$  \qquad  $C_{2} < -\frac{C_{1}^2}{N}$  &stable focus      &2 elliptic sectors \vspace{2mm}\\
3.b & $C_{1} = 0$  \qquad  $C_{2} < 0$                   &centre            &2 elliptic sectors \vspace{2mm}\\
3.c & $C_{1} < 0$  \qquad  $C_{2} < -\frac{C_{1}^2}{N}$  &unstable focus    &2 elliptic sectors \vspace{2mm}\\
\hline
\hline 
\end{tabular}
\end{flushleft}
\caption{Types of fixed points for linear system (LS) and the topology of trajectories for
nonlinear system (NLS). Definitions: $C = C_{1}^2 + N C_{2}$, where $N = 4$ for linear system and 
$N =2$ for nonlinear system.}
\label{types}
\end{table}

\begin{figure}
\begin{tabular}{ccc}
\vspace{-6mm} \\
\hspace{-6mm} \includegraphics[angle=90,width=60 mm,height=60mm]{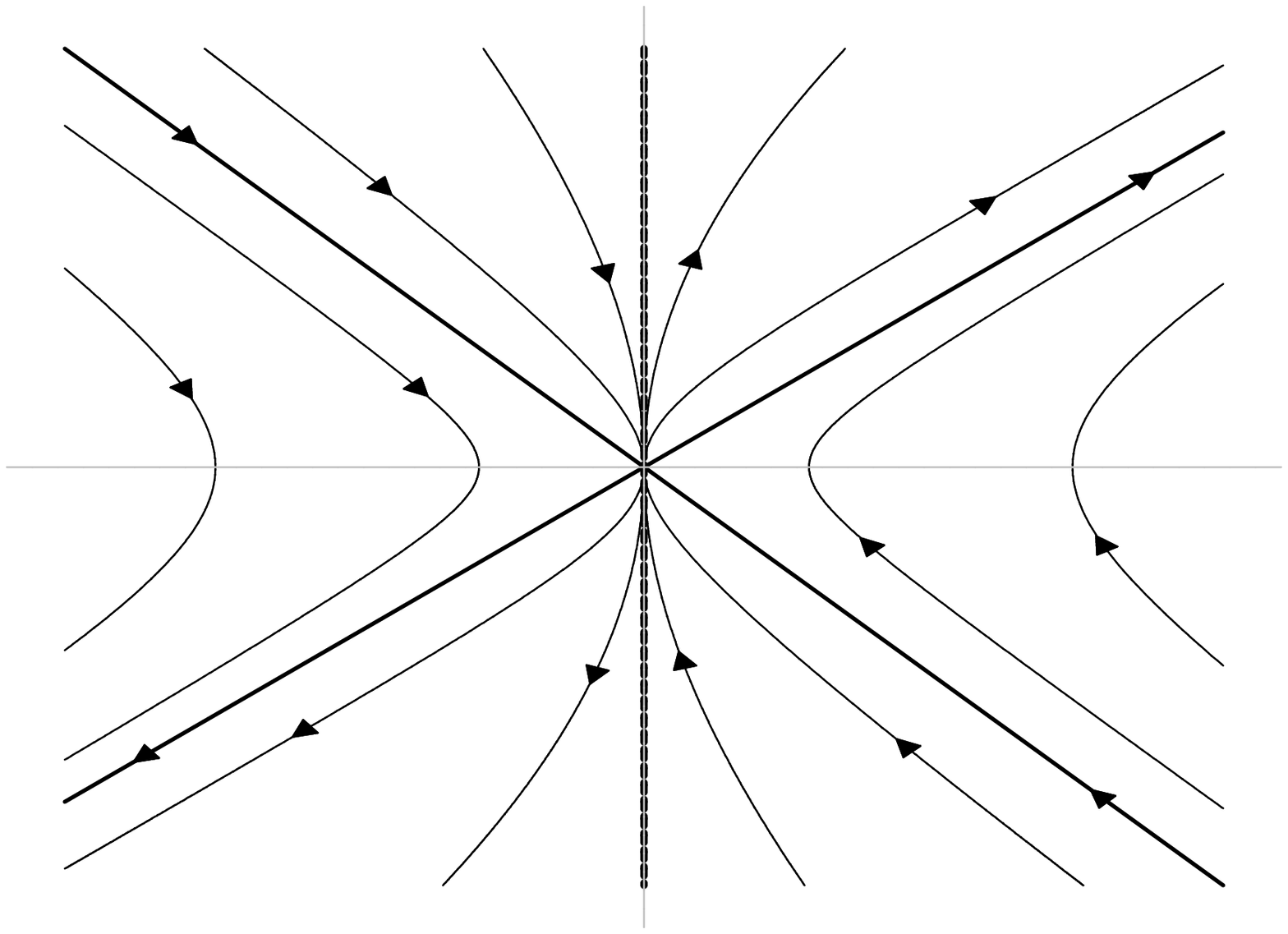} \hspace{-6mm} & 
\hspace{-6mm} \includegraphics[angle=90,width=60 mm,height=60mm]{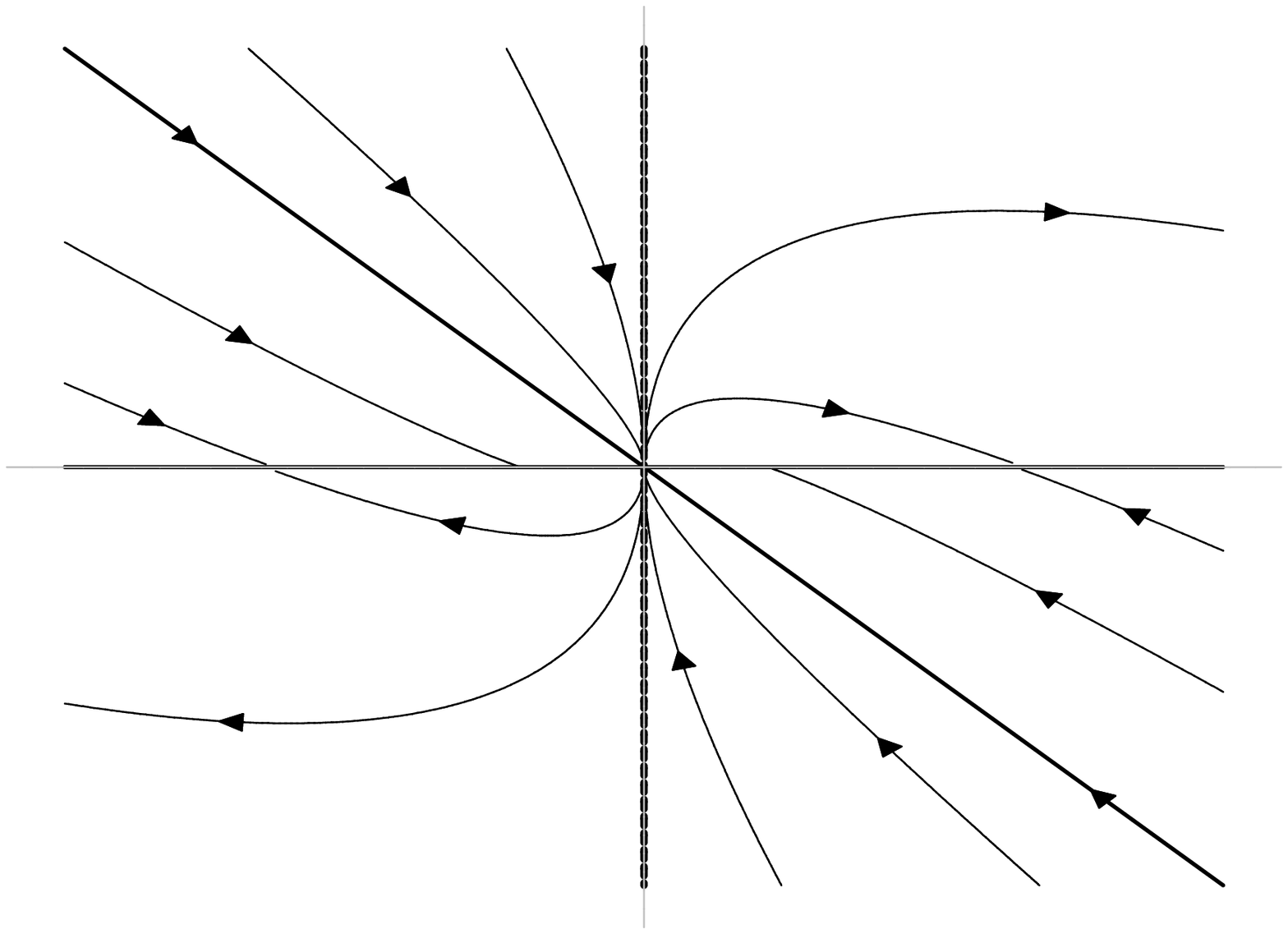} \hspace{-6mm} & 
\hspace{-6mm} \includegraphics[angle=90,width=60 mm,height=60mm]{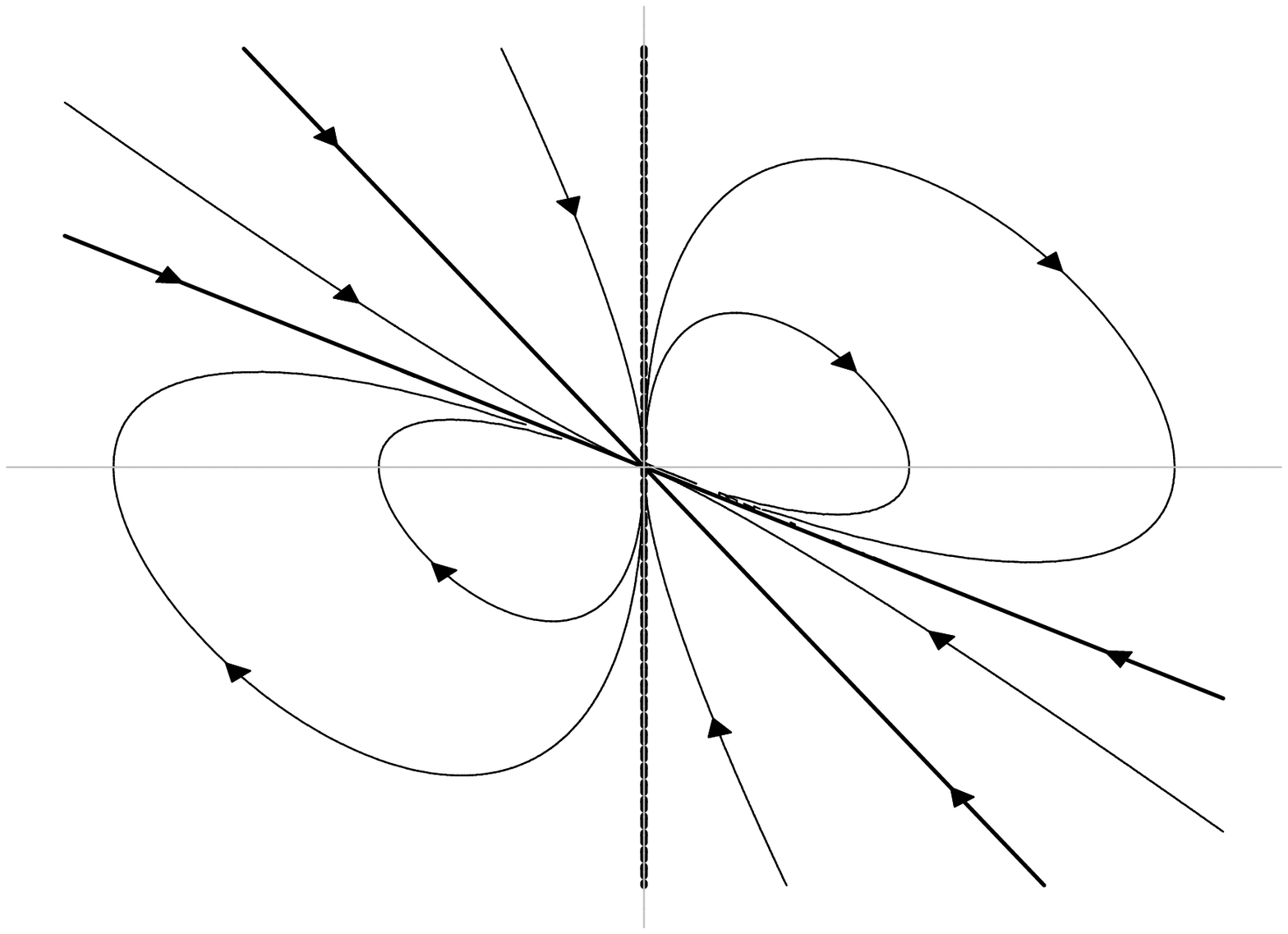} \hspace{-6mm}  
\vspace{-6mm} \\
\hspace{-6mm} 1.a (1.d) \hspace{-6mm} & 
\hspace{-6mm} 1.b       \hspace{-6mm} & 
\hspace{-6mm} 1.c       \hspace{-6mm} 
\vspace{-6mm} \\
\hspace{-6mm} \includegraphics[angle=90,width=60 mm,height=60mm]{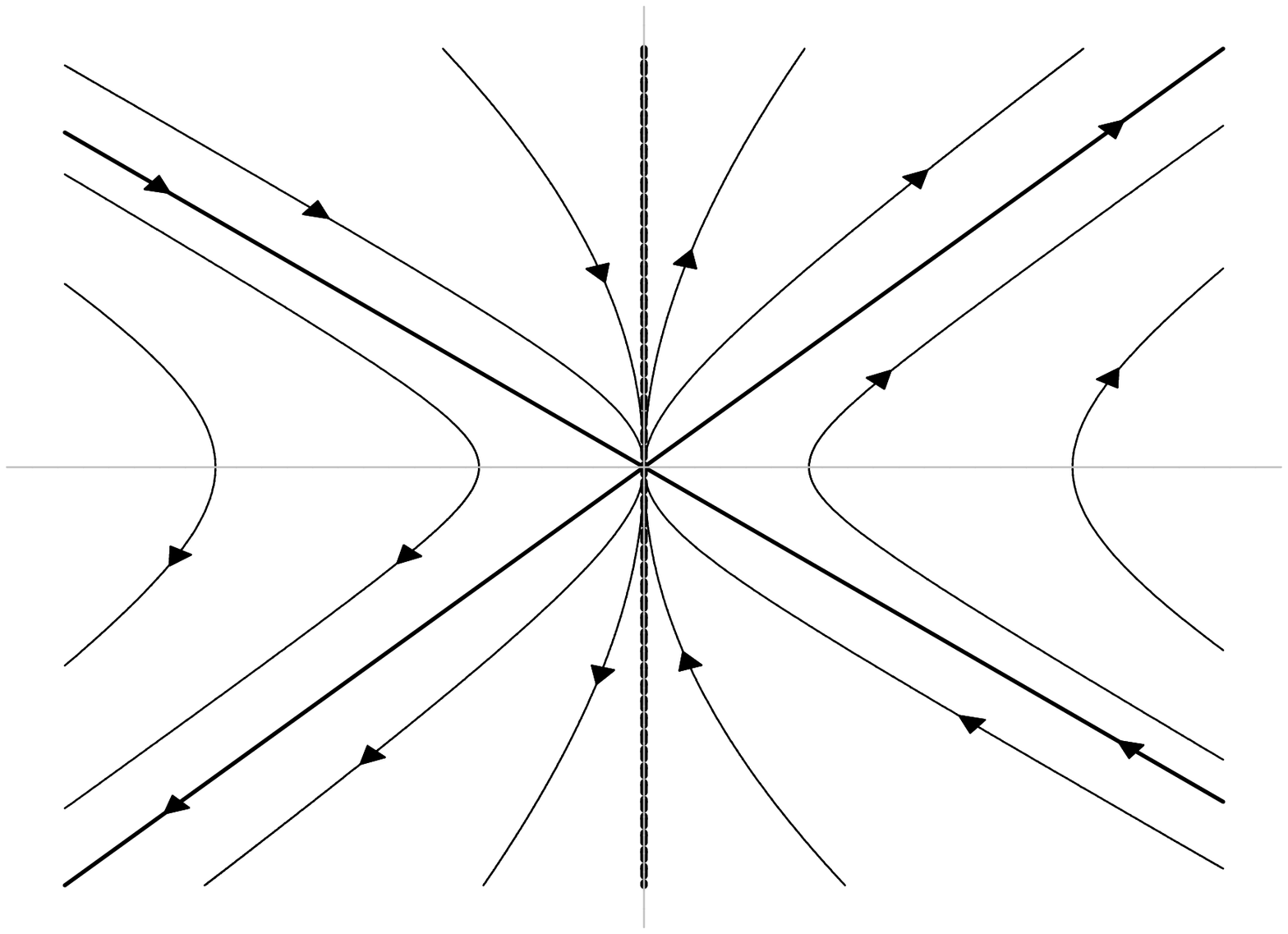} \hspace{-6mm} & 
\hspace{-6mm} \includegraphics[angle=90,width=60 mm,height=60mm]{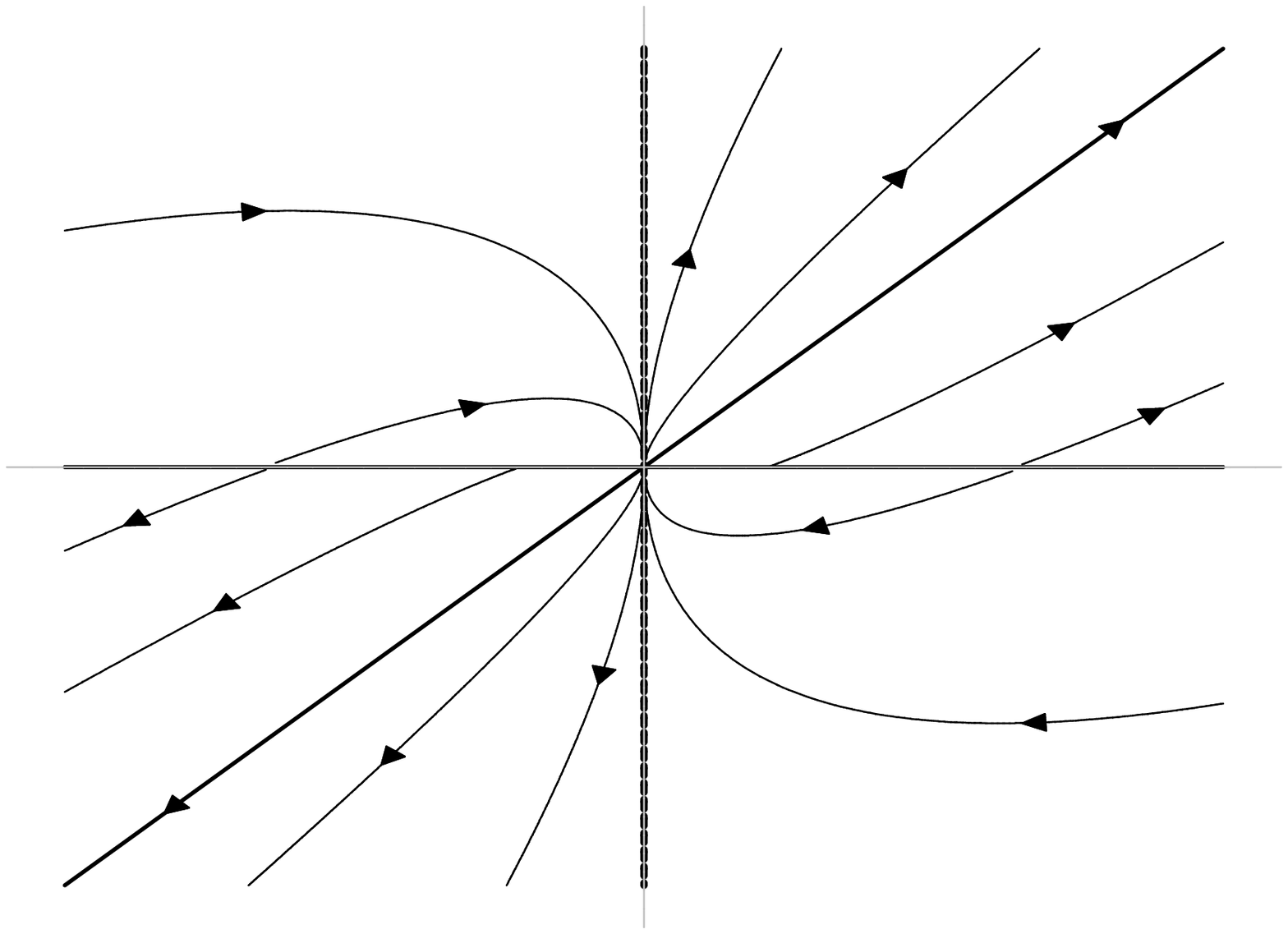} \hspace{-6mm} & 
\hspace{-6mm} \includegraphics[angle=90,width=60 mm,height=60mm]{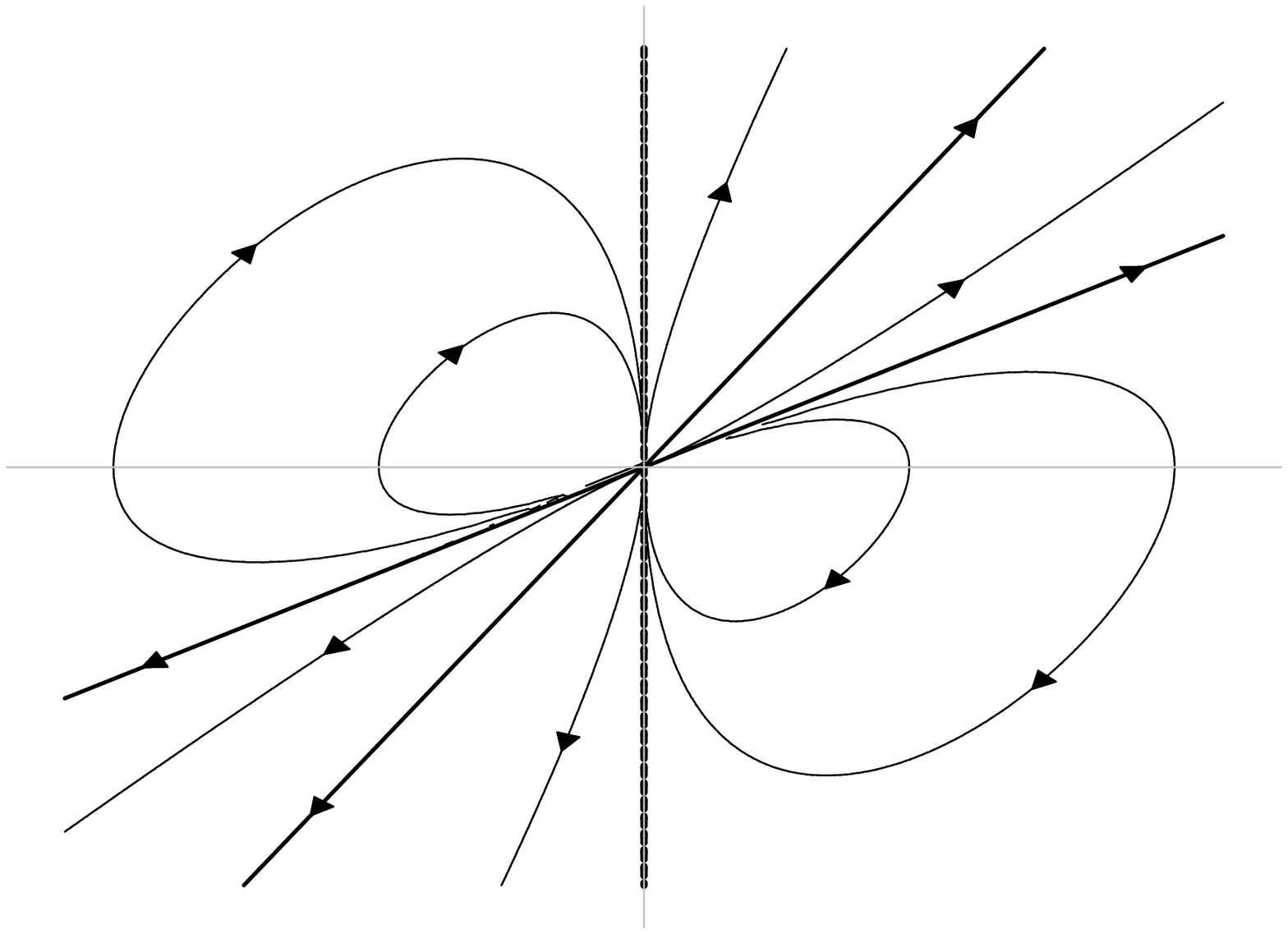} \hspace{-6mm}  
\vspace{-6mm} \\
\hspace{-6mm} 1.e (1.d) \hspace{-6mm} & 
\hspace{-6mm} 1.f       \hspace{-6mm} & 
\hspace{-6mm} 1.g       \hspace{-6mm} 
\vspace{-6mm} \\
\hspace{-6mm} \includegraphics[angle=90,width=60 mm,height=60mm]{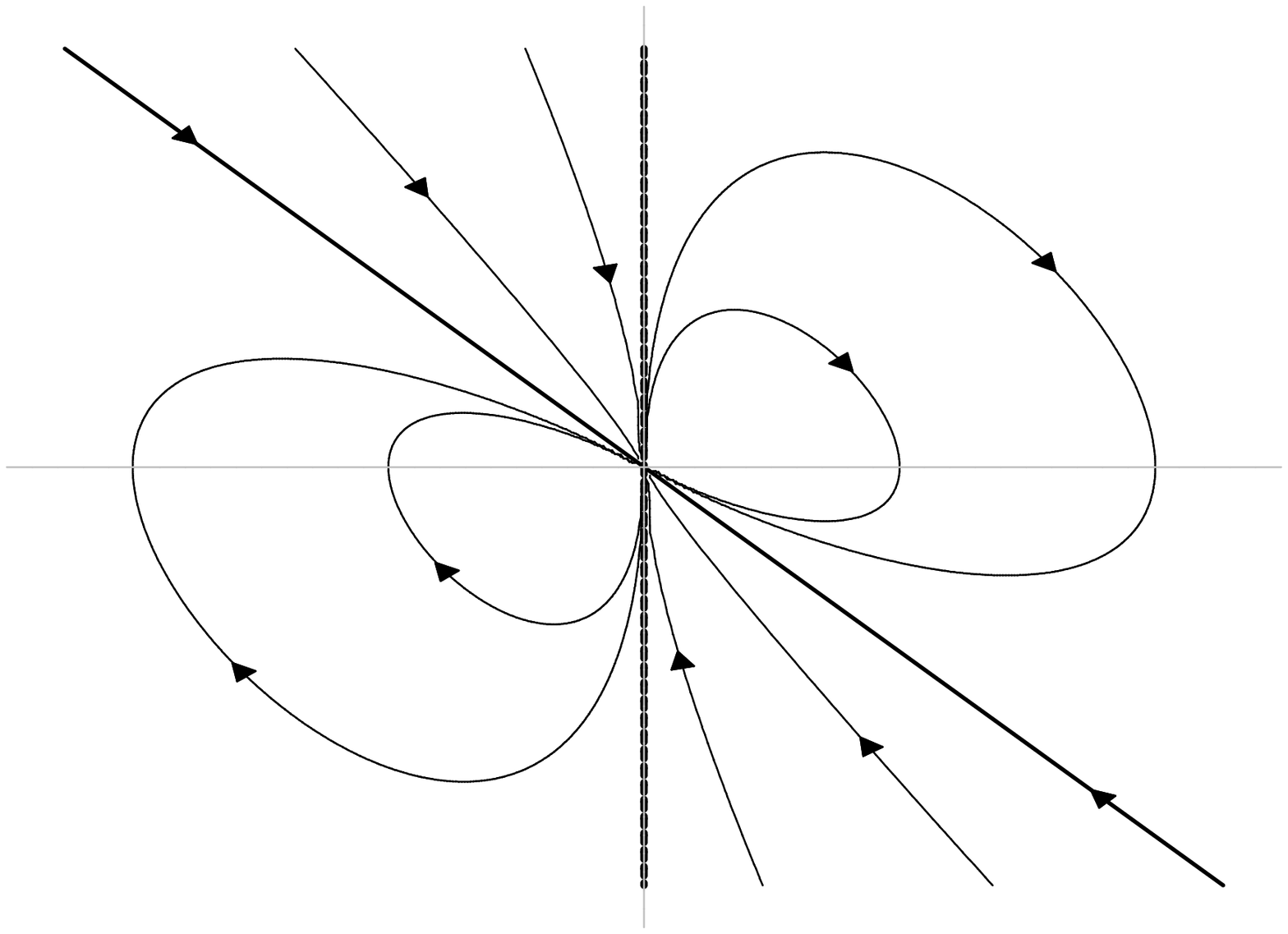} \hspace{-6mm} & 
\hspace{-6mm} \includegraphics[angle=90,width=60 mm,height=60mm]{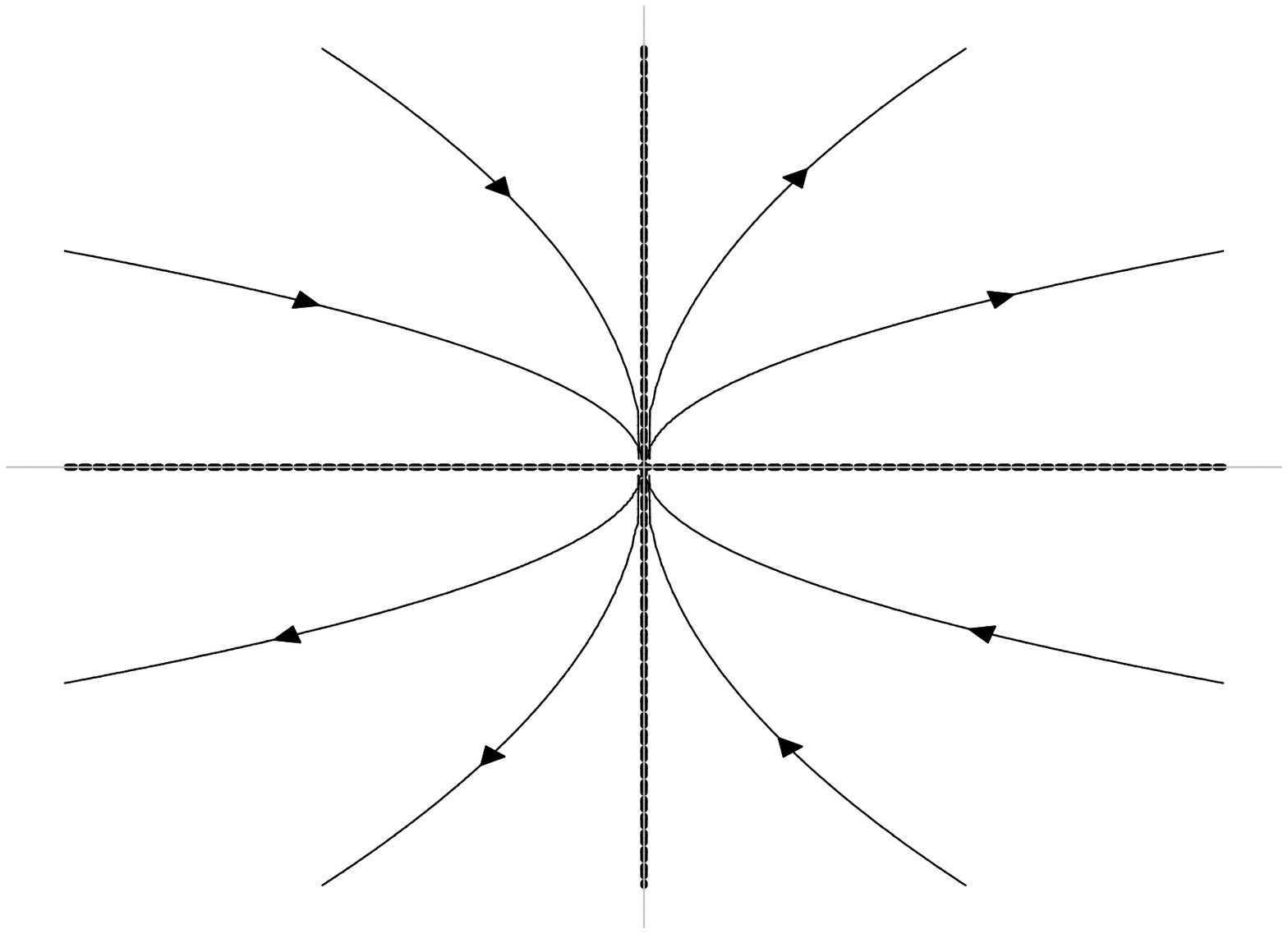} \hspace{-6mm} & 
\hspace{-6mm} \includegraphics[angle=90,width=60 mm,height=60mm]{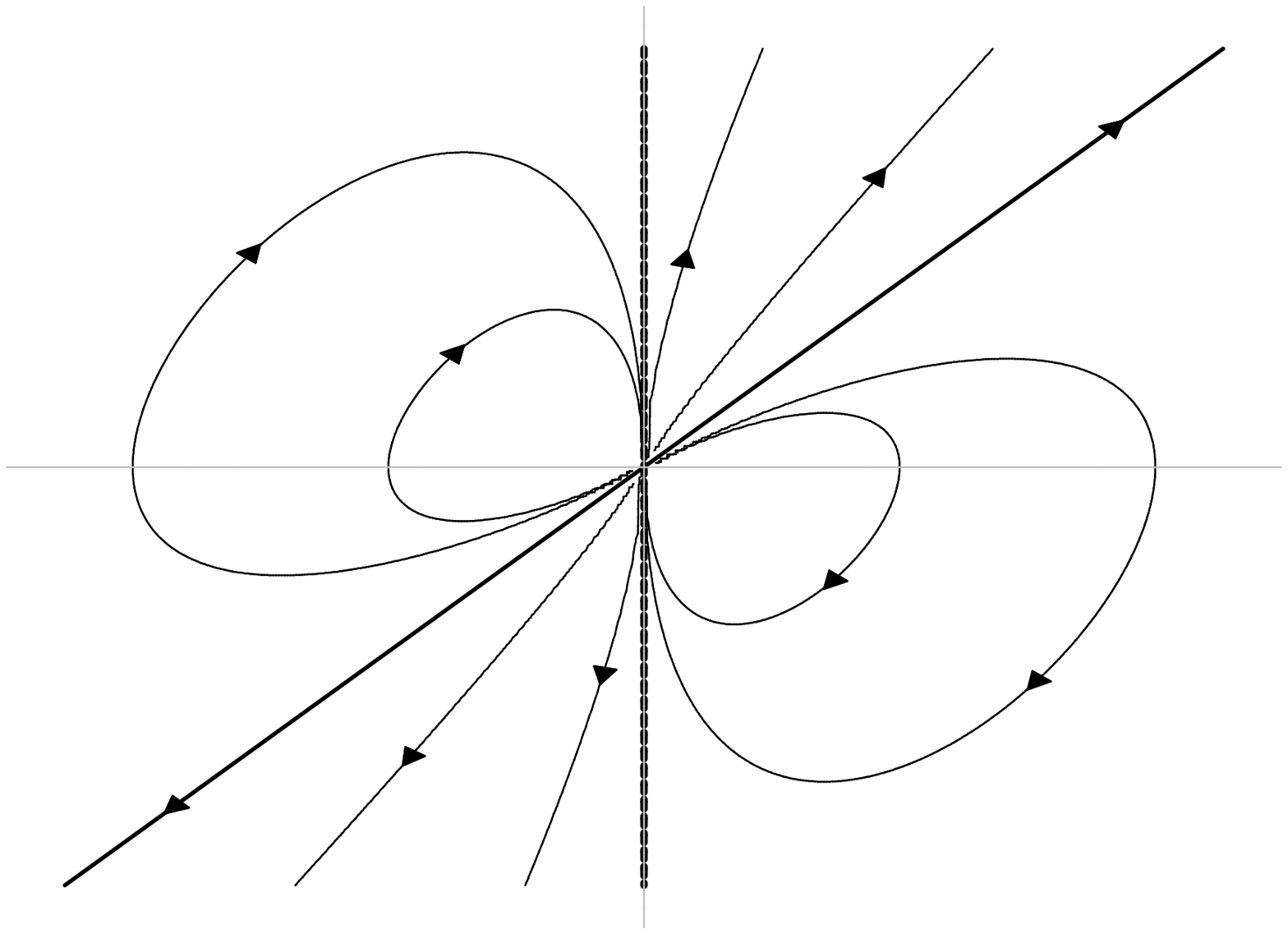} \hspace{-6mm} 
\vspace{-6mm} \\
\hspace{-6mm} 2.a  \hspace{-6mm} & 
\hspace{-6mm} 2.b  \hspace{-6mm} & 
\hspace{-6mm} 2.c  \hspace{-6mm} 
\vspace{-6mm} \\
\hspace{-6mm} \includegraphics[angle=90,width=60 mm,height=60mm]{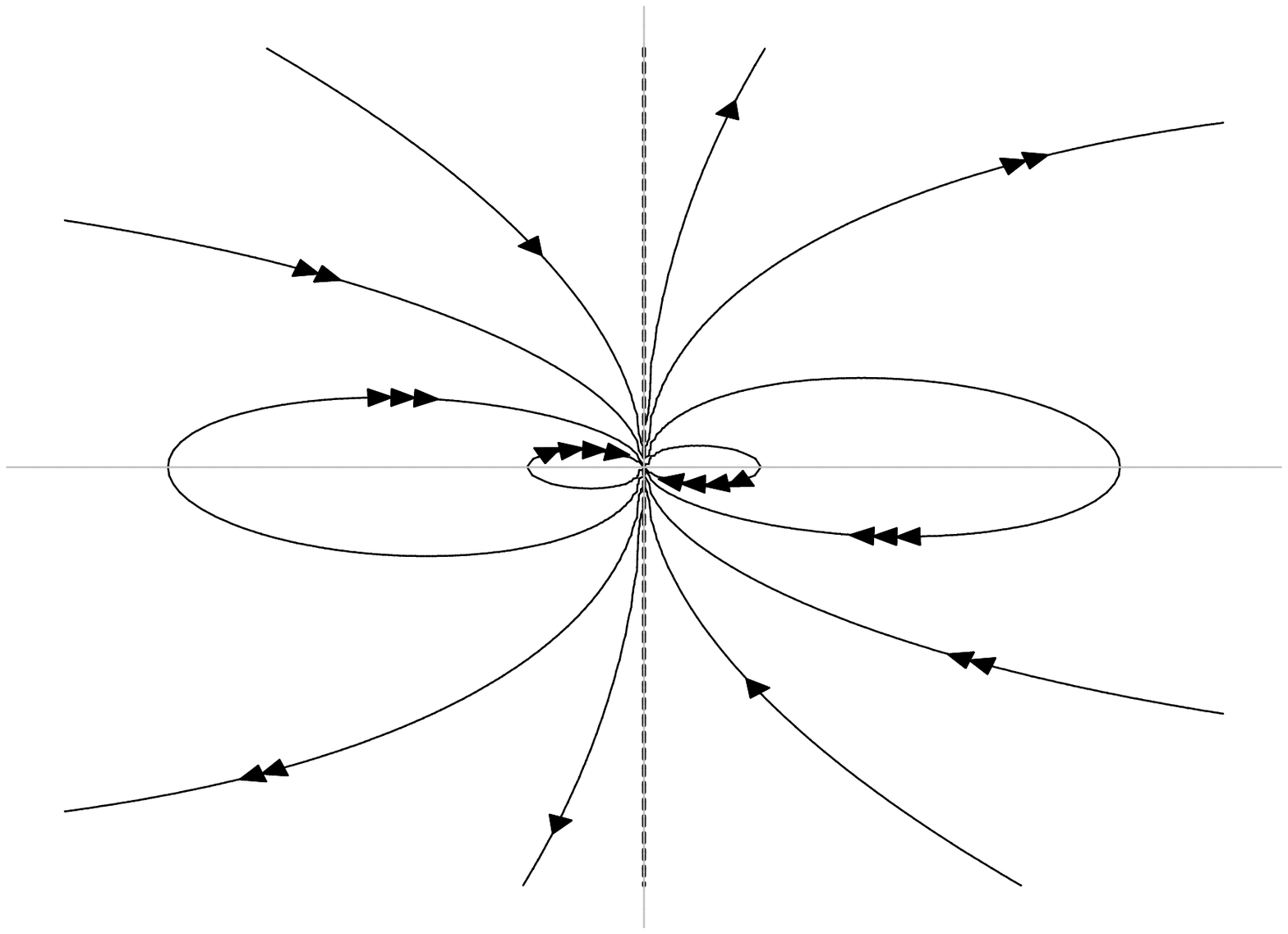} \hspace{-6mm} &
\hspace{-6mm} \includegraphics[angle=90,width=60 mm,height=60mm]{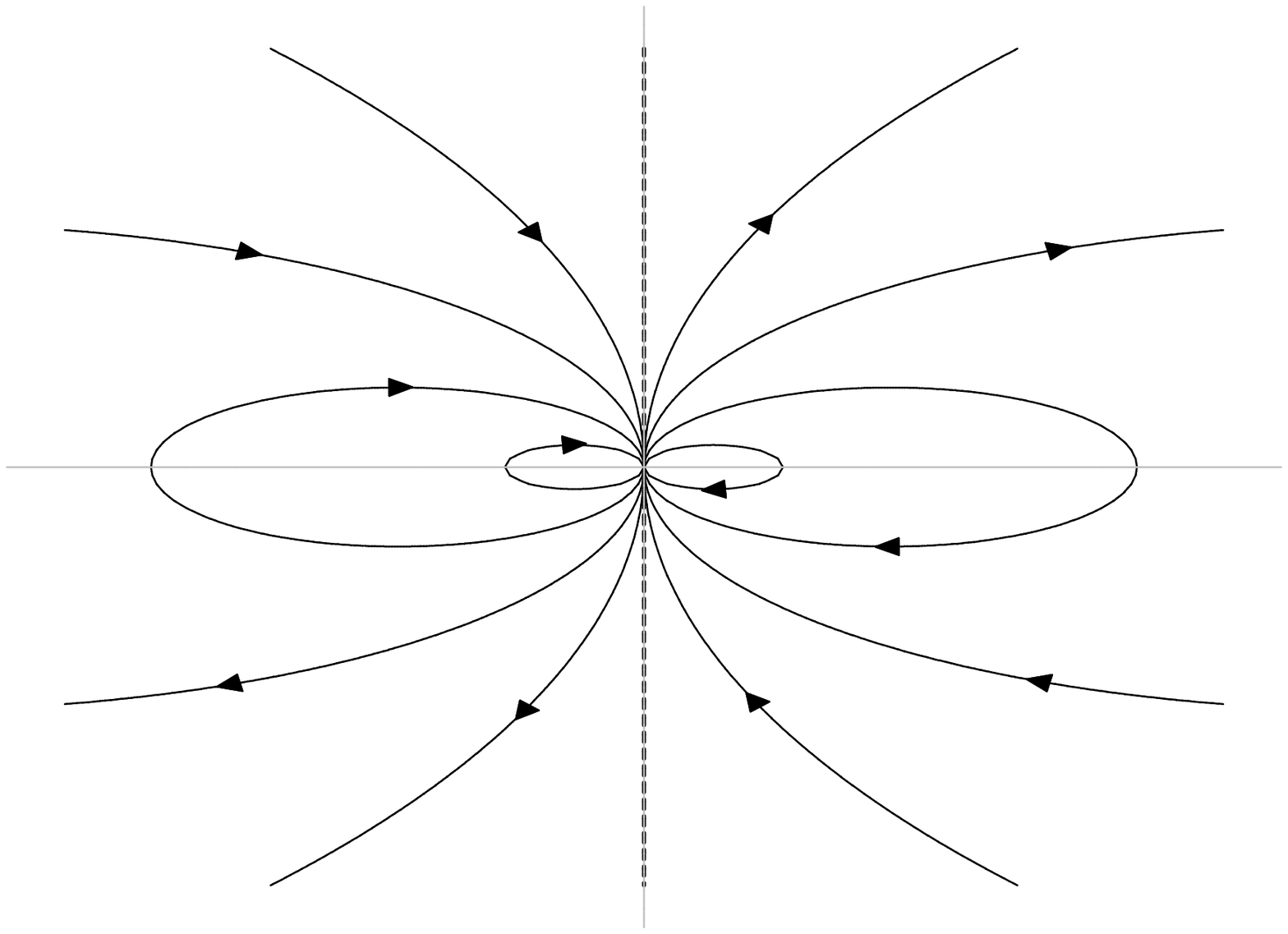} \hspace{-6mm} &
\hspace{-6mm} \includegraphics[angle=90,width=60 mm,height=60mm]{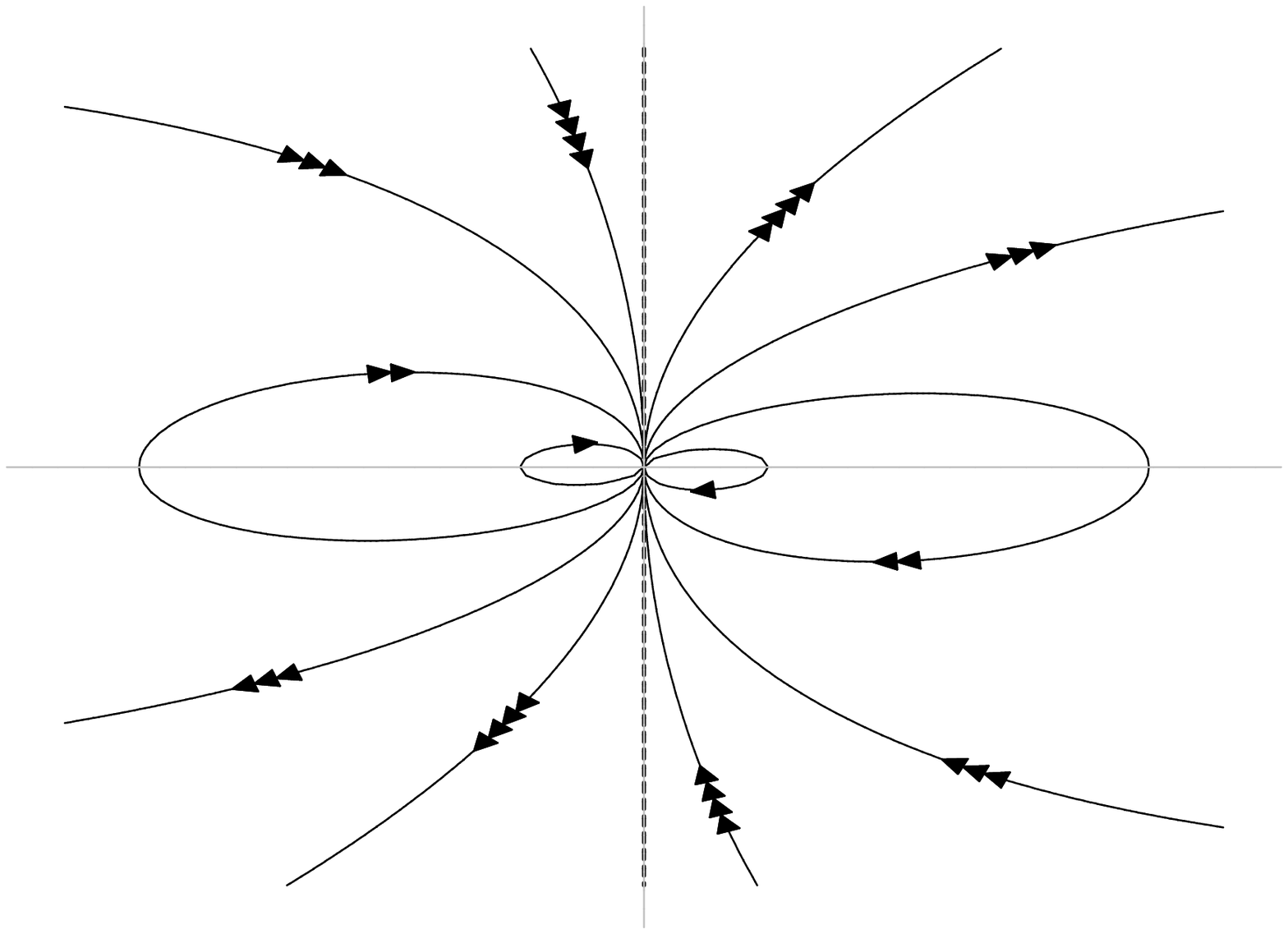} \hspace{-6mm}  
\vspace{-6mm} \\
\hspace{-6mm} 3.a  \hspace{-6mm} & 
\hspace{-6mm} 3.b  \hspace{-6mm} & 
\hspace{-6mm} 3.c  \hspace{-6mm} \\
\end{tabular} 
\caption{Phase portraits of the nonlinear approximation (\ref{eq_orb_ml}) near the GR point.}
\end{figure}
\footskip = 2.0 cm

\section{A special example} \label{trivial}

To further illustrate how well the full system and its approximations near the GR point 
fit together, let us consider a STG with a specific coupling function and potential,
\be
\omega (\Psi) = \frac{3 \Psi}{2(1-\Psi)} \qquad \qquad V(\Psi) = 0 \,.
\ee
From the condition $2\omega +3 \geq 0$ it follows that $\Psi \leq 1$. The 
GR point is at $\Psi_{\star} = 1$, $\Pi_{\star} = 0$, while the rest of
the border line, $\Psi_{\star} = 1$, $\Pi_{\star} \neq 0$, is deemed to be outside of the definition of the system.

The full equation for phase trajectories, (\ref{orbit_ex}), reads now
\be 
\frac{d \Pi}{d \Psi} = -\frac{3 \Pi}{2 \Psi} \big(-1 \pm \frac{1}{\sqrt{1 - \Psi}} \big) - \frac{\Pi}{2 (1 - \Psi)} 
\ee
Fortunately it is amenable to integration, the solution is
\be \label{triv_ex}
|\Pi| = K_1 \sqrt{1-\Psi}( 1 \pm \sqrt{1 - \Psi})^3  \,,
\ee
where $K_1$ is a constant of integration. 

Zooming to the vicinity of the GR point by (\ref{expans}), 
$1 -\Psi = x >0 $, $\Pi = y$, the solution (\ref{triv_ex}) approximates to a
parabola $|y| = K_1 \sqrt{x}$, 
which by identifying the integration constants $K_1^2 = 2 K$ is in perfect accord
with the solution $y^2 = 2 K x$ 
of the nonlinear approximate equation 
(\ref{eq_orb_ml}) with $C_1 = 0$, $C_2 = 0$
(case 2.b). 
This provides further evidence 
that the behavior of the full dynamical system (\ref{dynsys_x}), (\ref{dynsys_y})
near the GR point is adequately approximated by the nonlinear system (\ref{mlin_x}), (\ref{mlin_y}). 
Note also, that in the linear approximation we get a free motion, 
i.e. straight lines ${\tilde y} = \sqrt{\tilde{K}}$, which qualitatively gives 
a correct asymptotic state for the trajectories (flow away from the GR point), but clearly
does not provide a 
faithful phase portrait in comparison with the full solution.


\section{Summary and Discussion} \label{disc}

This paper considers general scalar-tensor gravity (STG) in the Jordan frame with arbitrary 
coupling $\omega(\Psi)$ and potential $V(\Psi)$.
We have presented and justified an approximate theory for the behavior of the scalar field 
in flat FLRW cosmological STG models in a potential dominated era and 
in the regime where the local weak field experiments are satisfied.
In terms of the phase space ($\Psi$, $\dot{\Psi} \equiv \Pi$) the latter is understood as the 
neighbourhood of the `GR point' ($\Psi_{\star}$, $\Pi_{\star}$),
defined by (a) $\frac{1}{2 \omega(\Psi_{\star}) +3} = 0$, (b) $\Pi_{\star}=0$.
We propose that if (c)  $\frac{d}{d \Psi}(\frac{1}{2 \omega(\Psi_{\star})+3}) \neq 0$ 
and (d) the higher derivatives of $\frac{1}{2 \omega(\Psi_{\star})+3}$ do not diverge,
then in the neighbourhood of the GR point
the nonlinear system (\ref{mlin_x}), (\ref{mlin_y})  
can be considered as an adequate approximate description of 
the full dynamical system (\ref{dynsys_x}), (\ref{dynsys_y}),
since both are endowed with the same characteristic features.
The phase portraits, summarized in Table 1 and depicted on Figure 1, 
typically show many trajectories passing through the GR point
either once on multiple times. In the end, only if 
\begin{displaymath}
V(\Psi_{\star})>0 \,,  \qquad \quad
\frac{d}{d \Psi}\left( \frac{1}{2 \omega(\Psi)+3} \right)\Bigg|_{\Psi_{\star}} \left(2V(\Psi)-\frac{dV(\Psi)}{d\Psi}\Psi \right)
\Bigg|_{\Psi_{\star}} <0 \,
\end{displaymath}
does the GR point function as an asymptotic attractor for the flow of all trajectories
in the vicinity. 

These analytic results could not have been predicted by numerical simulations,  
as the numerical calculations become rather problematic near the GR point 
due to the indeterminacy present
in the equations.
It would be very intersting to study how the different looping behaviors through the GR point
manifest themselves in terms of observational predictions.
It would also be of obvious physical relevance to extend the analysis to the matter dominated case,
although the treatment of the problem would face a difficulty of
having an additional phase space dimension to deal with.




\bigskip
{\bf Acknowledgments}
\smallskip

This work was supported by the Estonian Science Foundation Grant No. 7185 and by 
Estonian Ministry for Education and Science Support Grant No. SF0180013s07. 
M.S. also acknowledges the Estonian Science Foundation Postdoctorial research Grant No. JD131.

\end{document}